# Synthetic dimension band structures on a Si CMOS photonic platform

Armandas Balčytis[1], Tomoki Ozawa[2], Yasutomo Ota[3,4], Satoshi Iwamoto[3,5,6] Jun Maeda[1], Toshihiko Baba[1]

[1]Department of Electrical and Computer Engineering, Yokohama National University,79-5 Tokiwadai, Hodogaya-ku, Yokohama 240-8501, Japan
[2]Advanced Institute for Materials Research, Tohoku University, Sendai 980-8577, Japan
[3]Institute for Nano Quantum Information Electronics, The University of Tokyo, 4-6-1 Komaba, Meguro-ku, Tokyo 153-8505, Japan
[4]Department of Applied Physics and Physico-Informatics, Keio University, 3-14-1 Hiyoshi, Kohoku-ku, Yokohama 223-8522, Japan
[5]Research Center for Advanced Science and Technology, The University of Tokyo, 4-6-1 Komaba, Meguro-ku, Tokyo 153-8904 Japan
[6]Institute of Industrial Science, The University of Tokyo, 4-6-1 Komaba, Meguro-ku, Tokyo 153-8505, Japan
armandas.balcytis@gmail.com

**Abstract:** Synthetic dimensions, which simulate spatial coordinates using non-spatial degrees of freedom, are drawing interest in topological science and other fields for modelling higher-dimensional phenomena on simple structures. We present the first realization of a synthetic frequency dimension on a silicon ring resonator photonic device fabricated using a CMOS process. We confirm that its coupled modes correspond to a 1D tight-binding model through acquisition of up to 280 GHz bandwidth optical frequency comb-like spectra, and by measuring the first synthetic band structures on an integrated device. Furthermore, we realized two types of gauge potentials along the frequency dimension, and probed their effects through the associated band structures. An electric field analogue was produced via modulation detuning, whereas effective magnetic fields were induced using synchronized nearest- and second-nearest-neighbor coupling. Creation of coupled mode lattices and two effective forces on a monolithic Si CMOS device represents a key step towards wider adoption of topological principles.

The concept of dimensionality has become a central fixture in diverse fields of contemporary physics and technology in past years. While inquiries into lower-dimensional materials and structures have been fruitful, rapid advances in topology science have uncovered a further abundance of potentially useful phenomena beyond the three spatial dimensions available in the world around us. Recently, the concept of a synthetic dimension has emerged, with which high-dimensional structures can be hosted on low-dimensional platforms by harnessing non-spatial degrees of freedom as either a substitute or a complement for geometrical ones [1]. Owing to the flexibility in tuning both the amplitude and the phase of inter-site couplings along the synthetic dimension, the approach has become an important ingredient in realizing various topological lattice models ever since it was pioneered in ultracold atomic gases [2, 3].

Optics and photonics have also become fields in which simulation of various condensed matter phenomena and topological effects is drawing considerable attention [4]. This interest is owed to the versatility of photonics as a powerful platform for investigation of questions in fundamental science, as well as to the promise of novel applications that topological effects hold for the realization of next generation photonic devices. Examples of their remarkable functionalities and properties resilient to imperfections include robust transmission lines [5, 6, 7, 8], reflection-free sharply bent waveguides [9, 10, 11], or stable laser cavities [12, 13]. Here too, synthetic dimensions have made it possible to exploit higher-dimensional concepts in lower-dimensional devices with reduced complexity, as well as driving critical device functionalities such as on-chip optical isolation [14]. Harnessing states along the internal degrees of freedom of a photon as a substitute for or a supplement to a geometrical dimension creates a way of accessing advanced phenomena such as realizing photonic topological insulators [15], higher-order multipole moments [16], Anderson localization [17], higher dimension quantum Hall effect [14], and photonic Weyl points [18].

Past realizations of photonic synthetic dimensions include orbital angular momentum, polarization, delay between pulses and, in particular, frequency of light [19]. In constructing a synthetic space, optical frequency is an attractive variable as it can be practically partitioned into a lattice by utilizing conveniently equidistant ring resonator modes, as well as due to its relevance to important applications in frequency translation and spectral shaping of light. Very recently the first experimental realization of a synthetic frequency dimension was demonstrated using a macroscale optical fiber loop [20] and later a synthetic Hall ladder was realized by harnessing the direction of propagation within a ring as an additional dimension [21]. However, to enable robust platforms for researching complex multidimensional effects and creating practical optical devices, it is vital to miniaturize and transfer photonic realizations of synthetic dimensions into a monolithically integrated format. Early efforts towards this goal involved modelling thin-film lithium niobate electro-optic frequency combs as a multi-dimensional tight-binding lattice [22]. However, utilizing a more scalable silicon photonic chip platform would provide a considerable advancement, as it on one hand would allow photonics with synthetic dimensions to benefit from the mature and sophisticated CMOS commercial fabrication toolbox, and on the other, create the means for multi-dimensional topological phenomena to be introduced into novel device applications.

We report the realization of a synthetic frequency dimension on a Si CMOS platform by employing a custom dynamically modulated integrated ring resonator cavity design. By driving the resonator at its free-spectral range (FSR) rate of $\Omega_R$ = 20.4 GHz a frequency lattice spanning a 280 GHz bandwidth was established and sideband intensity enhancement was observed. The equidistant ring modes and periodic modulation of the cavity respectively represent lattice sites and their coupling mechanism, hence could be mapped to a 1D tight-binding model. By detecting the time resolved transmittance during modulation at the FSR rate we were able to experimentally measure the band structure involving a synthetic frequency dimension in an integrated photonic device for the first time. Furthermore, we demonstrate an ability to engineer inter-site couplings and induce non-reciprocity through photonic gauge potentials. By respectively employing FSR-detuned modulation frequencies or multiple modulation signals we were able to attain time-resolved demonstrations of band structures indicative of behaviors analogous to effective electric and magnetic field effects for photons. We suggest that such CMOS process fabricated ring resonator devices – supporting an extensive frequency dimension with versatile and reconfigurable coupling mechanisms, as well as two synthetic forces – can act as robust building blocks for scalable integrated circuits with nontrivial topology, which are promising for both answering fundamental questions in physics and realizing novel device prototypes.

## Results

**Device design.** Our approach to produce a lattice of coupled states along a synthetic frequency dimension involves the use of a ring resonator [20, 22]. Provided the waveguiding medium comprising the cavity exhibits near-zero

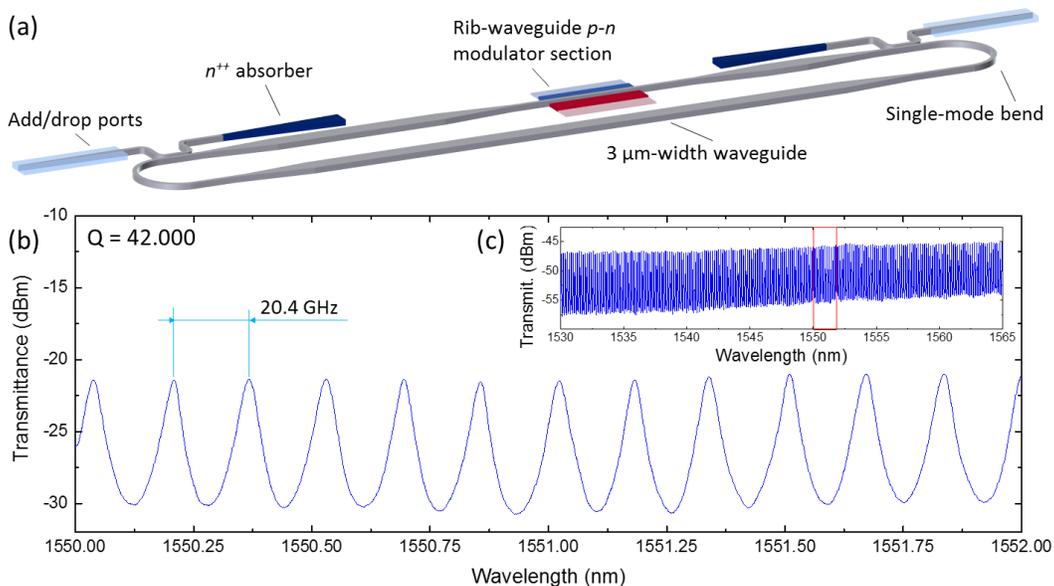

Fig. 1 (a) Schematic of the custom modulator-equipped ring resonator cavity design for the implementation of a synthetic frequency dimension. (b) Detailed drop-port transmittance spectrum in the λ = 1550 nm region spanning a 12 FSR interval with an approximate 20.4 GHz mode spacing, highlighted in inset (c) survey spectrum of the device throughout the telecommunication C-band.

group velocity dispersion in the relevant spectral region, it supports a spectrally equidistant sequence of states inherently akin to a 1D lattice. Coupling between different states can be induced using appropriately selected perturbation mechanisms, which for photonic devices can be generalized as either extrinsic modulation, which is the approach chosen in this work, or self-induced nonlinear effects [19]. Lastly, artificial electric and magnetic fields for shaping photon flow can be produced through modulation of the coupling constants between different states along a dimension, be it synthetic [20, 23] or conventional [24].

A schematic of our Si CMOS platform-compatible ring resonator equipped with an in-cavity phase modulator is provided in Fig. 1(a). It was designed to have an approximately 20 GHz FSR that could be bridged with reasonable efficiency using electronic modulation at radio frequencies (RF), thereby ensuring coupling between adjacent modes. This necessitates a total cavity circumference of $l \approx 4.1$ mm, which is a significant reduction in form-factor compared to prior demonstrations using 13.5 m length optical fiber loops [20]. Management of optical propagation losses is a key aspect of the device, as observation of frequency lattice effects is contingent on photon decay lifetime being maximized relative to the lattice hopping rate. To mitigate internal loss in the resonator, the majority of its passive optical path was established in a 3-µm-wide multimode, operating as quasi-singlemode, waveguide. Singlemode 400-nm-wide waveguides were nevertheless employed where optical confinement was needed, such as for waveguide bends and input/output coupling sections. Cavity modulation was applied using a p-n doped rib waveguide phase shifter segment, which was limited to a 155 µm length in order to minimize losses. Optical input and output for the device was performed in the add/drop configuration in the weak coupling condition.

**Cavity characterization.** Ring resonators were fabricated by employing commercial CMOS foundry services using a 200-mm-diameter silicon-on-insulator wafer process. Experimental transmittance spectra, plotted in Fig. 1(b), revealed that ring resonators exhibited equidistant mode spacing of 20.4 GHz, close to the specified 20 GHz FSR value, and an approximately 10 dB extinction spectrum owing to the weak coupling regime. Furthermore, as depicted in Fig. 1(c), the spectral distribution remained notably consistent throughout the telecommunication C-

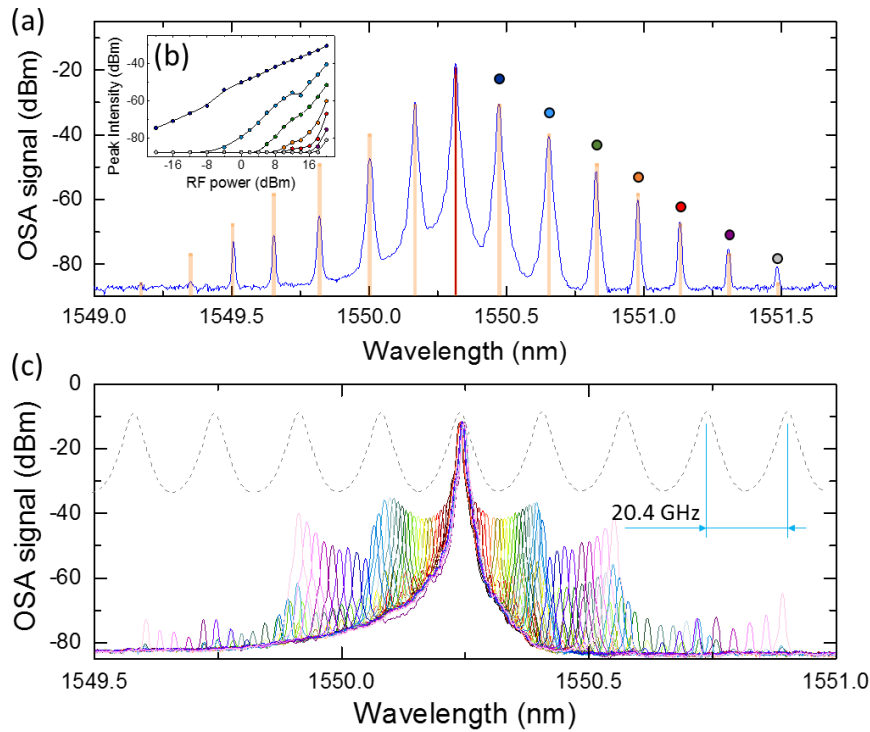

Fig. 2 (a) Optical output spectrum of a ring cavity under on-resonance 10 dBm optical excitation and FSR-matched 20.4 GHz frequency and 20 dBm power RF modulation. Orange bars represent an effective 1D tight-binding model with a 1.46 loss/coupling strength ratio. (b) Dependence of sideband peak spectral power density on modulating RF power. (c) Overlapping modulation spectra as 15 dBm power RF signal is swept over the 0 – 40 GHz range. The dashed line illustrates the ring cavity transmittance spectral function.

band wavelength region. Around the $\lambda$ = 1550 nm wavelength region the loaded cavity possessed a quality factor of Q = 42,000 that corresponds to a photon decay rate of $2\gamma$ = 4.5 GHz, with dissipation primarily attributable to insertion losses in the p-n modulator section.

For a ring resonator to operate as a lattice along a frequency dimension, a coupling mechanism for photons to translate between different modes has to be induced through an external time-dependent perturbation of the dielectric properties of the cavity at a rate equal to (or a fraction of) an optical round-trip period [14]. The most straightforward way of achieving this involves dynamical modulation of the ring resonator at its FSR frequency. In practice this means ensuring that modulation sidebands overlap with adjacent ring modes. Optical output spectrum in Fig. 2(a) illustrates an optimized case for our device, in which one of its resonances is pumped (spectral position is indicated by the red line) while the cavity is continuously driven by a FSR-matched $\Omega_R$ = 20.4 GHz RF signal.

Modulated device spectra reveal that the mode $m$ under external laser excitation couples to the first nearest FSR modes ($m \pm 1$), but this coupling also cascades to modes further down the lattice ($m \pm 2$, $m \pm 3$ …). It is evident that cavity losses are sufficiently low for cascaded frequency conversion to generate comb lines at a notable distance from the pump laser wavelength. The Si ring resonator device proved capable of hosting a long synthetic frequency dimension of up to 14 coupled modes spanning a 280 GHz bandwidth, as can be seen in Fig. 2(b). This span is determined by the bandwidth over which the innate waveguide material dispersion allows for constructive interference needed for cascaded frequency conversion [25], as well as in-cavity optical losses.

A direct demonstration of electrically induced coupling between adjacent modes is provided in Fig. 2(c), where overlapping optical output spectra are plotted as modulation frequency is swept from 0 GHz to 40 GHz in 1 GHz increments. As the modulation sidebands approach a 20.4 GHz FSR spacing they become resonantly enhanced [26], and this increase of their spectral intensity follows the ring cavity transmittance function. Furthermore, the modulation enhancement response was significant at 2×FSR 40.8 GHz and 3×FSR 61.2 GHz RF frequencies, indicating that the electrical bandwidth of the p-n junction phase shifter segment is sufficiently high to enable up to third-nearest-neighbor mode coupling.

**Mapping to the 1D tight-binding model.** Our device can be interpreted as a tight-binding lattice model, in which particles localized at specific sites can hop to neighboring ones at a certain rate. The model Hamiltonian which simulates the behavior of our system in the presence of nearest- and next-nearest-neighbor hopping is:

$$H = \sum_m \varepsilon_m \hat{b}_m^\dagger \hat{b}_m - \sum_m (J_1 \hat{b}_m^\dagger \hat{b}_{m+1} + e^{i\phi} J_2\, \hat{b}_m^\dagger \hat{b}_{m+2} + H.c.), \qquad (1)$$

where $\hat{b}_m^\dagger$ denotes a bosonic particle creation operator at a site indexed by an integer $m$, $\varepsilon_m$ represents the position-dependent on-site potential and $J_1$ and $e^{i\phi} J_2$ are the nearest- and next-nearest-neighbor hopping amplitudes. We allowed the next-nearest-neighbor hopping to take a complex value with phase $\phi$ which will later allow us to tune the effective magnetic field present in the synthetic ladder. The position dependence of the on-site potential, $\varepsilon_m$, on the other hand, will allow us to introduce an effective electric field.

A realistic situation in a photonic system has to consider optical driving and continual energy leakage [14]. Such losses can be taken into consideration by way of a mode-dependent loss rate $\gamma_m$, and they are balanced by the continuous injection of monochromatic light at a rate $f_m(t) = f_m e^{-i\omega t}$. It has been demonstrated that the cavity field expectation value $\beta_m(t) \equiv \langle \hat{b}_m(t) \rangle e^{-i\omega t}$ evolves according to the equation of motion [27]:

$$i \frac{\partial}{\partial t} \beta_m = (\varepsilon_m - \omega)\, \beta_m - J_1 \beta_{m+1} - J_1 \beta_{m-1} - e^{i\phi} J_2\, \beta_{m+2} - e^{-i\phi} J_2\, \beta_{m-2} - i\gamma_m \beta_m + f_m\,. \qquad (2)$$

Assuming a steady state, Eq. (2) can adequately simulate the frequency synthetic dimension lattice in Fig. 2(a), and the result corresponding to a $\gamma/J_1 \approx 1.46$ loss to coupling strength ratio condition has been plotted against the experimental spectrum using orange bars. Deviations from symmetry in the experimentally measured optical output are attributable to influence of effective refractive index dispersion in the Si ring cavity.

One of the primary advantages of employing the synthetic frequency dimension approach for lattice simulations is that, since inter-mode coupling is induced via an externally applied perturbation mechanism, various models can be realized with relative ease by altering the applied modulation pattern. Optical output spectra in Fig. 3(a) and (b) show the steady state response of the microring device under RF modulations of the form $V(t) = V_0\cos(\Omega t)$ at frequencies $\Omega$ = 20.4 and $2\Omega$ = 40.8 GHz, resonantly matched respectively one or two FSR spans. As illustrated in the inset sketches, these two cases correspond to different tight-binding lattice coupling configurations. Fig. 3(a) relates to a basic nearest-neighbor coupled tight binding model, whereas Fig. 3(b) shows the equivalent of a next-nearest neighbor coupled tight binding model in which the available states are divided between two 1D sub-lattices (only one of which is optically pumped). This, in addition to the ability to tune the coupling strength by changing modulation voltage, provides remarkable versatility to synthetic dimension devices compared to geometrical space resonant cavity array counterparts, in which both coupling configuration and strength would be fixed during fabrication.

**Band structure measurements.** Experimentally measured band structures for the synthetic 1D tight-binding model with nearest-neighbor coupling at different modulation strength conditions are shown in Fig. 3(c). They were acquired using the method pioneered by by Dutt *et al* [20] for macroscale fiber loop cavities. Since for a modulated ring resonator system modes spaced along the frequency axis comprise a direct space, its corresponding reciprocal quasi-momentum axis $k$ is equivalent to time. Here the Brillouin zone corresponds to a single $T = 50$ ps timescale optical round-trip inside the ring resonator device. When modulation frequency $\Omega$ is matched to the FSR frequency $\Omega_R$, RF oscillation period is likewise equal to the duration of the round-trip in a ring cavity. Laser detuning $\Delta\omega$ is equivalent to the energy axis, and here it is scanned over a 3 FSR range revealing the same number of Floquet bands. The dispersion relation for a 1D tight-binding model in case of resonant modulation is given by the simple periodic function:

$$\varepsilon_k = 2J_\eta \cos(k\eta\Omega_R), \tag{3}$$

where $J_\eta$ is the mode coupling strength over distance $\eta$, $k$ is the quasi-momentum, and $\Omega_R$ is the modulation frequency. Fits of this dependence are overlaid onto the experimental band structures in Fig. 3 and show good agreement. The nearest-neighbor mode coupling strength parameter derived from this fitting could be set from zero (flat bands) up to $J_1 = 0.864$ GHz when maximal 20 dBm RF power was employed. While the value of $J_1$ was somewhat lower than the $2\gamma = 4.5$ GHz loss rate in the modulator-equipped Si ring cavity, evidenced by considerable width of the Floquet bands and characterized by a $\gamma/J_1 \approx 2.6$ loss to coupling strength ratio, it nevertheless permitted temporally resolved band structure observations.

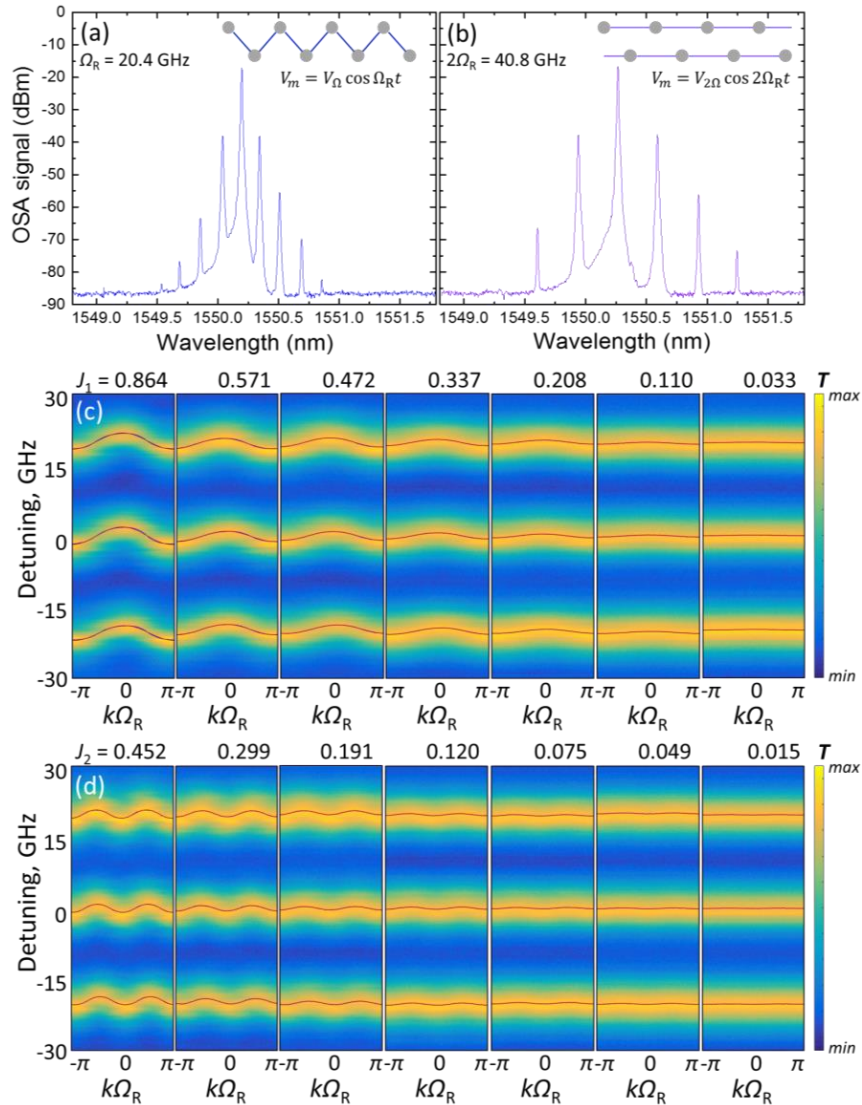

Fig. 3 Optical output spectra of a microring device under on-resonance FSR-matched (a) 20.4 GHz frequency 12 dBm power, and (b) 40.8 GHz frequency 20 dBm power RF modulation. Insets show sketches of corresponding tight-binding lattice models. Synthetic frequency dimension bands at different on-resonance modulation powers for (c) FSR 20.4 GHz and (d) 2×FSR 40.8 GHz. Rightmost is the weak modulation case, and the mode coupling strength $J_\eta$ [GHz] increases towards the left. Superimposed plots are the dispersion curve of 1D tight binding model, $\varepsilon_k = 2J_\eta \cos k\Omega$ fits.

In the case of a 1D tight-binding model with next-nearest adjacent mode coupling, dispersion relations, shown in Fig. 3(d), exhibit two modulation periods within an optical round-trip delimited Brillouin zone. As this configuration requires the application of an RF signal at a $\Omega = 40.8$ GHz spanning over a 2×FSR range, modulation efficiency was lower due to finite bandwidth of the device. However, at maximum RF power, 1D tight-binding dispersion relation fitting-derived coupling strength parameter was $J_2 = 0.452$ GHz and clearly resolvable in experiment. Similar measurements conducted using $\Omega = 61.2$ GHz 3×FSR modulation for $J_3$ were less conclusive due to experimental limitations.

**Demonstration of an effective electric field.** To fully exploit the synthetic dimension in integrated optical systems it is not sufficient to merely establish a coupled lattice model for photons. Another essential component is a way to control photon transport in given lattices. In condensed matter systems this can be achieved by application of gauge potentials in the form of electric and magnetic fields. While photons do not couple to such fields, an equivalent behavior can be produced by engineering the on-site energy of each site and the hopping phase the particles accumulate as they translate over a sequence of lattice sites either in real or synthetic space [14].

A prominent example of an effective gauge potential in a synthetic dimension involves the induction of Bloch oscillations [28]. For a frequency lattice it requires merely selecting the modulation frequency $\Omega$ to be slightly yet appreciably different from the resonant mode spacing $\Omega_R$ (all experimental data presented up to this point operated at the $\Omega = \Omega_R$ condition). When modulation frequency is detuned by $\Delta\Omega$ so that $\Omega = \Omega_R - \Delta\Omega$, keeping in mind the equivalence between $k$ and $t$, Eq. (3) can be rewritten:

$$\varepsilon_k(t) = 2J_1\cos(k[\Omega_R - \Delta\Omega]) = 2J_1\cos(k\Omega_R - \Delta\Omega \cdot t). \tag{4}$$

This means that a ring resonator cavity modulated at a frequency shifted from an FSR value by a small magnitude is analogous to synthetic frequency dimension 1D tight binding lattice under a constant force, as sketched in Fig. 4(a). Its band structure, while still with a Brillouin zone set by one optical cavity round-trip, drifts over time along the $k$ axis at a rate proportional to the frequency detuning magnitude $\Delta\Omega$, as illustrated in Fig. 4(b). In experimental practice this means that the band structure acquisitions become dependent on laser wavelength scanning rate, hence can be considered dynamic. Very recently the first experimental demonstration of such dynamic acquisitions using a 10 m circumference optical fiber loop setup was reported by Li *et al* [23].

The dynamic band structures acquired on our integrated Si ring resonator under detuned modulation conditions are provided in Fig. 4 panels (c) and (d) respectively. Compared to the static case $\Omega = \Omega_R$ in Fig. 4(b), detuning introduces a time-dependent phase component proportional to frequency difference $\Delta\Omega$. It manifests as trajectories produced by the interplay of band structure drift along the quasi-momentum axis due to a constant synthetic force, and the rate at which the energy axis is scanned. Furthermore, these trajectories become inverted along the $k$ axis when modulation offset $\Delta\Omega$ sign is reversed. At $\Delta\Omega = \pm 100$ Hz, corresponding to a situation where $\Omega$ and $\Omega_R$ frequencies are close enough that their beating rate is comparable to experimental time resolved 2D transmittance spectral acquisition duration available in our setup, band structures merely appear distorted due to continuous drift. However, at $\Delta\Omega = \pm 300$ Hz and beyond, the drift becomes rapid enough for several Brillouin zone periods to be covered during a single acquisition and band splitting becomes apparent. While the fine structure of individual split bands is somewhat obscured by the considerable ring resonance linewidths in our device, simulated trajectories in Fig. 4(c) show a good match between experiment and theory. Therefore, this experiment exhibited all the key behaviors expected in a 1D tight-binding model lattice system under constant force.

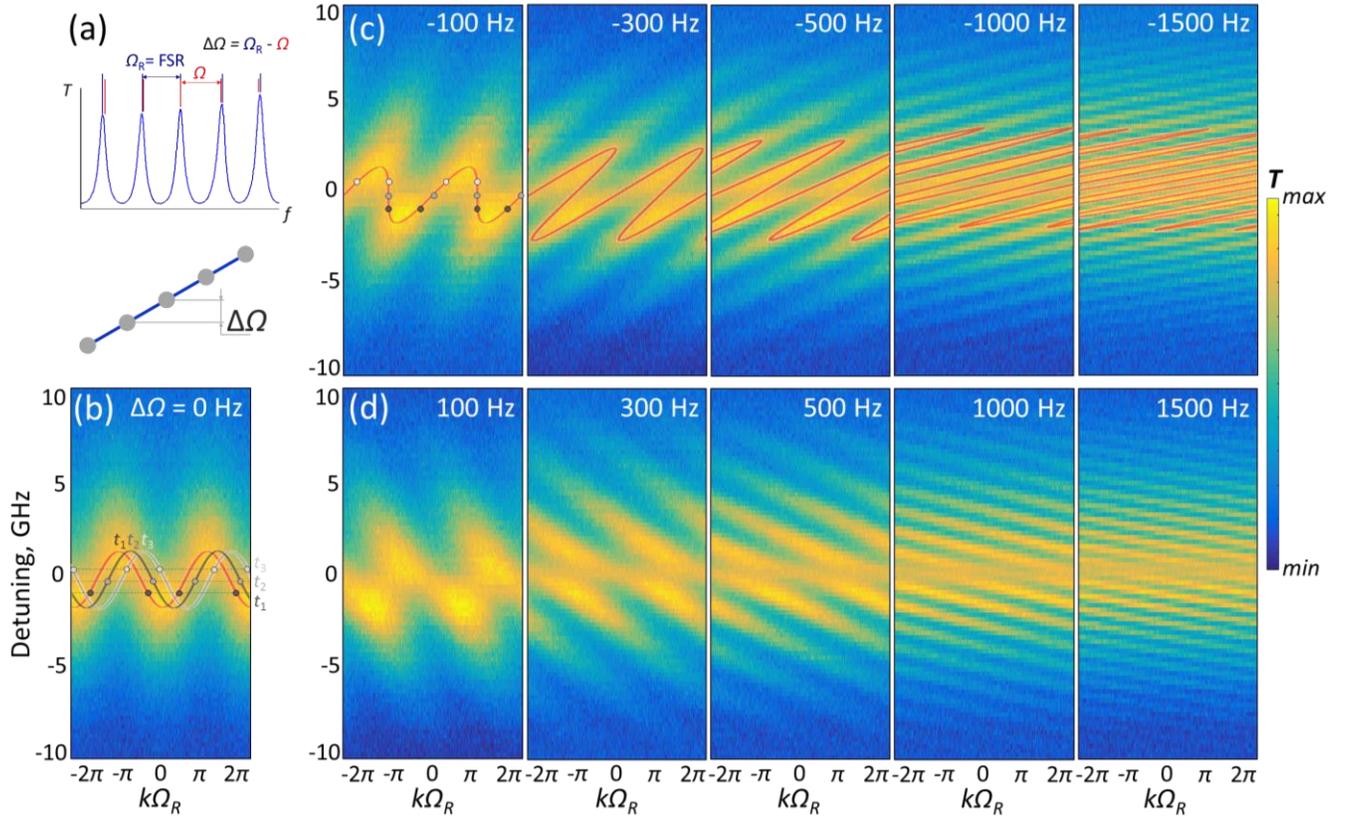

Fig. 4 Experimental measurement of dynamic band structures. (a) Illustration of the relationship between ring resonator spectrum, resonant optical round-trip FSR frequency $\Omega_R$ and the modulation frequency $\Omega$, shifted from resonance by $\Delta\Omega$. Diagram below shows how this system relates to a 1D tight-binding model of a photon under an effective force. (b) Static on-resonance band structure, spanning over two Brillouin zones for clarity. When modulation is off-resonance, bands gradually shift over time along the $k$ axis, so that transmittance acquisitions at different optical frequencies at times $t_1$, $t_2$, and $t_3$, generate a dynamic trajectory as plotted in (c) first panel. Experimentally measured band structures with different (c) negative and (d) positive modulation frequency detuning from $\Omega_R = 20.4$ GHz values. RF modulation power was $P_{RF} = 13$ dBm which produced an estimated $J_1 = 0.61$ GHz coupling strength. Trajectories drawn in red are guides for the eye.

**Demonstration of an effective magnetic field.** The photonic analogue of a magnetic field has been proposed in real space, which leads to an optical analog of the quantum spin Hall effect without time reversal symmetry breaking [29], as well as with temporal modulation by way of which this reciprocity is broken [24]. Feasibility of engineering such an effective gauge potential in a synthetic frequency space has been shown by applying dual-tone modulation with controllable strength and relative phase [20]. Realizing such a perturbation scheme in our integrated Si photonic device involves the simultaneous application of synchronized $\Omega_R = 20.4$ GHz and $2\Omega_R = 40.8$ GHz frequency signals, resulting in a modulation voltage of the form $V_m(t) = V_\Omega \cos(\Omega_R t + \varphi) + V_{2\Omega}\cos 2\Omega_R t$, where $\varphi$ is a phase offset introduced using a tunable RF delay component.

As illustrated in the top inset of Fig. 5, such dual-tone modulation induces simultaneous nearest- and second-nearest-neighbor 1D tight-binding lattice couplings. This partitions the ring resonator modes into forming a quasi-

2D triangular ladder lattice configuration with an alternating magnetic flux. An effective gauge potential effect can be observed through experimental band structures provided in Fig. 5, where a single FSR spectral span and one Brillouin zone are highlighted for clarity. Each of the depicted plots represent a different phase offset $\varphi$, hence a different effective gauge potential. Conversely, the modulation powers of $\Omega_R = 20.4$ GHz and $2\Omega_R = 40.8$ GHz RF signals where fixed to 12 dBm and 20 dBm to partly compensate for diminishing device response towards higher frequencies. Dispersion relations for this synthetic dimension 1D tight-binding model coupling configuration can be modelled using the equation:

$$\varepsilon_k = 2J_1\cos(k\Omega_R + \phi) + 2J_2\cos(k2\Omega_R), \tag{5}$$

where $J_1$ and $J_2$ are coupling strength parameters for adjacent and next-nearest modes respectively. Fitting of Eq. (5) to the experimental data, depicted as solid curves overlaying color plots in Fig. 5, shows a robust agreement between model and measurement. Based on the fits, aforementioned RF modulation powers yield coupling strength values $J_1 = 0.367$ GHz and $J_2 = 0.260$ GHz, resulting in a $J_2/J_1 \approx 0.6$ ratio.

The key feature of the synthetic approach is the flexibility afforded to band structure engineering through control of both coupling strength and gauge potential $\varphi$. We observed that, when $\varphi \neq 0, \pi$, the band structure is no longer symmetric about $k = 0$. This asymmetric band structure is a clear signature of time-reversal symmetry breaking [14, 20]. As the relative phase approaches $\varphi \approx \pi$ this non-reciprocity is removed, and the band structure becomes equivalent to the $\varphi \approx 0$ case with a $\pi$ Brillouin zone shift. The system ultimately returns to its initial state at $\varphi = 2\pi$. We thereby demonstrate how full control over a magnetic field analogous gauge potential can be achieved in an integrated Si photonic device.

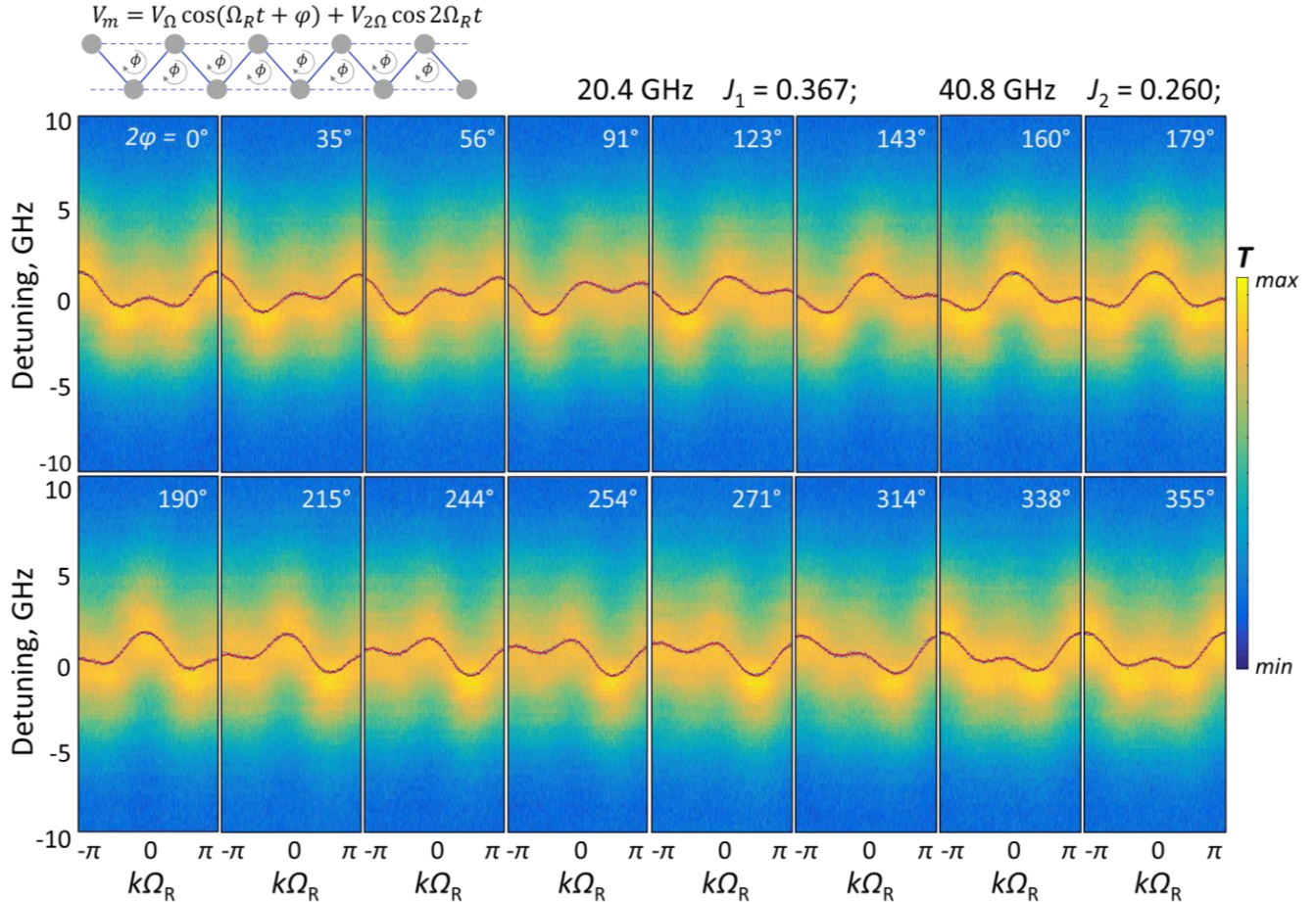

Fig. 5 Band-structure engineering using nearest-neighbor coupling (20.4 GHz), long-range hopping (40.8 GHz) and a synthetic magnetic field ($\varphi$) when coupling strength ratio $J_2/J_1 \approx 0.6$ is maintained constant. Variation of 20.4 GHz modulation phase delay $\varphi$ induces gradual band structure changes. The phase delay increases from left to right. Superimposed blue and red plots are the Lorentzian center frequency and $\varepsilon_k = 2J_2 \cdot \cos k2\Omega_R + 2J_1 \cdot \cos(k\Omega_R + \varphi)$ fits, respectively, and inset illustrates the applicable quasi-2D tight-binding model.

**Discussion**

We presented and experimentally characterized the first realization of an integrated synthetic frequency dimension device on a Si CMOS platform. The ring resonator, equipped with an in-cavity phase shifter, exhibited a $\Omega_R = 20.4$ GHz FSR and was capable of supporting a 1D tight-binding model-equivalent lattice of up to 14 coupled states spanning a 280 GHz bandwidth. Further confirmation of this model's applicability was made by way of time resolved band structure acquisitions. In addition, two types of photonic gauge potentials, namely equivalents of electric and magnetic forces, have been demonstrated on our Si device platform through their effects on the aforementioned band structures. This work opens the path to harnessing the mature and highly scalable silicon integrated photonics fabrication infrastructure to tackle pressing challenges in quantum simulation and topology science. The high integration density attainable on the Si CMOS platform is poised to greatly increase the level of complexity of lattice models that can be realized on a chip, and to even pursue novel applications for designing optical devices such as isolators [14], or enacting arbitrary linear transformations of photon frequencies [30] by using integrated cavities analogous to the one presented here as a foundational building block.

Despite a long list of advantages, which includes high integration density, low cost, and a leading level of standardization, the most notable challenge of using a silicon device platform is optical loss that limits the range of observable emergent lattice phenomena through photon decay. The monolithically integrated device presented here has losses that are among the lowest achievable for a contemporary CMOS process, with wide quasi-singlemode waveguides exhibiting transmittance similar to those achieved using etchless fabrication approaches [31]. However, some further optimization is possible with respect to phase shifter design, for instance, by harnessing silicon-polymer hybrid modulator technology [32]. Recent advances in hybrid-integrated semiconductor optical amplifier components show significant promise as a method for compensating in-cavity losses [33]. Furthermore, optical gain would enable the investigation of non-Hermitian topology in both real and synthetic dimensions [34] and greatly expand the wealth of accessible physical phenomena. Some restrictions on the attainable lattice models are also due to limited bandwidth of available measurement equipment, with up to approximately 65 GHz frequency signal generators and photodetectors capable of inducing up to third or fourth-adjacent neighbor coupling between resonant modes for rings with practical FSRs.

Promising avenues for further investigation include the creation of more complex integrated device lattice models in the synthetic frequency dimension, such as a Su–Schrieffer–Heeger configuration [16], or creating an effectively 2D tight-binding lattice in a modulated 1D array of rings [14]. Harnessing gauge potentials in the frequency dimension is particularly promising, as it would pave the way for the creation of novel frequency converters [28]. Conversely, development of the synthetic dimension framework will be fruitful for further understanding of cascaded photonic phenomena, such as for rapid yet comprehensive characterization of optical frequency comb signals [35]. The inherent flexibility and reconfigurability of the synthetic dimension approach is complementary with analogous static lattices in real space, and is expected to enable novel devices that circumvent the 2D constraints imposed by planar integrated photonics and that could model phenomena beyond three dimensions.

**Methods**

**Steady state characterization:** Transmittance spectra were acquired using a telecom wavelength Santec TSL-550 excitation laser source with up to 12 dBm power output, and an Advantest Q8221 optical multi power meter was used for drop port output detection. Light was coupled into and out of the silicon-on-insulator device edge coupler mode converters via lensed fibers. Optical spectra under RF modulation were detected using an Advantest Q8384 optical spectrum analyzer. RF driving of ring resonator devices was done using continuous wave sinusoidal signals with frequencies from 1 GHz to 61.2 GHz and output powers ranging from -20 dBm to 20 dBm, supplied by a Rohde & Schwarz SMA100B generator. The maximum optical excitation and RF modulation powers that can be applied are limited respectively to 10 dBm and 20 dBm by the onsets of nonlinear responses. Electronic interfacing with the modulator metal contact pads on the integrated device was conducted using a G-S-G 3-point RF probe.

**Band structure measurements:** Since the reciprocal space of a synthetic frequency dimension corresponds to time, measurement of dispersion relationships of such a system involves temporally resolved acquisitions of optical power in the ring cavity. This was done by monitoring the device drop port output using an Agilent Infinium DCA-J 86100C oscilloscope with an up to 50 GHz RF detector module. Prior to that, optical-to-electronic signal conversion was performed using an Anritsu MN4765B optical receiver module with an up to 65 GHz bandwidth. Optical excitation laser was a Santec TSL-550 device operating at a 10 dBm power and wavelength in the vicinity of $\lambda = 1550$ nm. Device output signal strength attainable for the carrier frequency was around −15 dBm and first sidebands were −9 dB below that. Therefore the optical signal was amplified using a low noise Alnair Labs LNA-220-C erbium doped fiber amplifier, which supplied up to 20 dB of amplification, and an Alnair Labs CVF-220CL tunable band pass filter (BPF) to suppress the amplified spontaneous emission noise. The BPF that was set to a moderately broad $\Delta\lambda_{BPF} = 2$ nm pass band to allow to scan the probing laser over a several FSR wavelength range without sideband clipping. The electronic portion of the setup is comprised of a set of RF generators phase locked using their internal 10 MHz clocks. In the basic 1D tight-binding model case, a Rohde & Schwarz SMA100B RF source was used for driving the ring resonator device, and a lower frequency Anritsu MG3692B generator supplied a $\Omega_R/10$ trigger to the oscilloscope. More complex simultaneous nearest- and next-nearest-mode coupling configurations were induced by employing additional phase locked HP 83640B RF generators, and their outputs where combined using a Midwest Microwave MFR 34078 up to 40 GHz bandwidth 3 dB power splitter/directional coupler. Relative phase of FSR and 2×FSR modulation waveforms could be adjusted using a PE8253 controllable delay element connected to the 20.4 GHz frequency input branch. The measurement proceeded by automatically scanning the input laser wavelength in $\Delta\lambda = 5$ pm increments and acquiring time resolved transmittance output oscilloscope traces over a set optical frequency detuning range. All of these traces are collected and arranged in sequence to compile a 2D transmittance plot, directly related to the synthetic frequency dimension band structure of the system.

**Acknowledgements**

The research is conducted as a JST CREST project Grant Number JPMJCR19T1. T. O. acknowledges support from JSPS KAKENHI Grant Number JP20H01845, JST PRESTO Grant Number JPMJPR19L2, and RIKEN iTHEMS. Armandas Balčytis is an International Research Fellow of JSPS. We thank Mr. T. Isomura, Nippon Sokki Co. Ltd., for help with the experiment.

**Author contributions**

A.B. designed and conducted device characterization and band structure acquisition experiments and performed data analysis. T.O. was responsible the theoretical modeling framework for device design and analysis. J.M. in consultation with Y.O., S.I. and T.B devised the device architecture. All authors contributed to discussion of the results and writing the manuscript. T.B. and S.I. supervised the project.

**Competing interests:** The authors declare no competing interests.